\documentclass[aps,prl,twocolumn,superscriptaddress,showpacs]{revtex4}
\usepackage{graphicx}
\usepackage{latexsym}
\usepackage{amssymb}
\usepackage{amsmath}
\usepackage{amsfonts}
\usepackage{bm}
\usepackage{multirow}
\usepackage{color}

\newcommand{\beq}{\begin{equation}}
\newcommand{\eeq}{\end{equation}}
\newcommand{\beqn}{\begin{eqnarray}}
\newcommand{\eeqn}{\end{eqnarray}}

\begin{document}

\title{Exotic Quantum Critical Point on the surface of $3d$ Topological Insulator}


\author{Zhen Bi}

\author{Yi-Zhuang You}

\author{Cenke Xu}

\affiliation{Department of physics, University of California,
Santa Barbara, CA 93106, USA}

\date{\today}

\begin{abstract}

In the last few years a lot of exotic and anomalous topological
phases were constructed by proliferating the vortex like
topological defects on the surface of the $3d$ topological
insulator
(TI)~\cite{TI_fidkowski1,TI_fidkowski2,TI_Qi,TI_senthil,TI_max}.
In this work, rather than considering topological phases at the
boundary, we will study quantum critical points driven by vortex
like topological defects. In general we will discuss a $(2+1)d$
quantum phase transition described by the following field theory:
$\mathcal{L} = \bar{\psi}\gamma_\mu (\partial_\mu - i a_\mu) \psi
+ |(\partial_\mu - i k a_\mu)\phi|^2 + r |\phi|^2 + g |\phi|^4$,
with tuning parameter $r$, arbitrary integer $k$, Dirac fermion
$\psi$ and complex scalar bosonic field $\phi$ which both couple
to the same $(2+1)d$ dynamical noncompact U(1) gauge field
$a_\mu$. The physical meaning of these quantities/fields will be
explained in the text. Making use of the new duality formalism
developed in
Ref.~\onlinecite{son,maxashvin,wangsenthil1,wangsenthil2}, we
demonstrate that this quantum critical point has a quasi self-dual
nature. And at this quantum critical point, various universal
quantities such as the electrical conductivity, and scaling
dimension of gauge invariant operators can be calculated
systematically through a $1/k^2$ expansion, based on the
observation that the limit $k \rightarrow + \infty$ corresponds to
an ordinary $3d$ XY transition.

\end{abstract}


\maketitle


{\it --- Introduction}

Although it is well-known that the boundary state of a
noninteracting $3d$ topological insulator (TI) is described by one
or odd number of free $(2+1)d$ Dirac
fermions~\cite{fukane,moorebalents2007,roy2008}, curiosity drives
theorists to look for all possible boundary states of $3d$ TI
under strong interaction. It was demonstrated that under strong
interaction the boundary of a $3d$ TI can have various topological
orders that cannot be realized in a pure $2d$
system~\cite{TI_fidkowski1,TI_fidkowski2,TI_Qi,TI_senthil,TI_max}.
And the general procedure of obtaining these topological orders,
is to first drive the boundary into the so-called Fu-Kane
superconductor~\cite{fukane2}, then restore the U(1) symmetry by
condensing a bosonic vortex of the superconductor, for example a
vortex of 8 fold vorticity (or a vortex that would trap $8
\frac{hc}{2e}$ flux once the fermion is coupled to the external
electromagnetic field). In the condensate of the $8-$fold vortex,
all symmetries of the system are preserved, the boundary remains
gapped, but the ground state has topological order with nonabelian
anyon
excitations~\cite{TI_fidkowski1,TI_fidkowski2,TI_Qi,TI_senthil,TI_max}.

More recent theoretical exploration has concluded that the charge
neutral $4-$fold vortex is a fermion, and it is doublet that
transforms under time-reversal symmetry as $\mathcal{T}: \psi
\rightarrow i\sigma^y \psi^\dagger$. This fermionic $4-$fold
vortex provides a dual description of the boundary of $3d$ TI,
which is a $(2+1)d$ quantum electrodynamics (QED$_3$) with $N=1$
flavor of Dirac fermion: \beqn \mathcal{L}_{dual} =
\bar{\psi}\gamma_\mu (
\partial_\mu - i a_\mu) \psi + \frac{1}{e^2} f_{\mu\nu}^2 , \cr\cr
\gamma^0 = \sigma^y, \ \ \gamma^1 = \sigma^x, \ \ \gamma^2 =
\sigma^z, \label{qed1}\eeqn where $a_\mu$ is the dual of the
Goldstone mode of the Fu-Kane superconductor, and the flux quantum
of $a_\mu$ carries half of the physical electric
charge~\cite{son,maxashvin,wangsenthil1,wangsenthil2}, thus
$a_\mu$ is a noncompact gauge field. This duality is a fermionic
version of the well-known duality between the $3d$ XY model and
the bosonic QED~\cite{halperindual,leedual}. And based on this
duality, recently it was demonstrated that QED$_3$ with $N = 2$ is
self-dual~\cite{xudual}, which is a fermionic analogue of the
self-duality of the noncompact CP$^1$ theory with easy-plane
anisotropy~\cite{ashvinlesik,deconfine1,deconfine2}.

Ref.~\onlinecite{maxashvin,wangsenthil1,wangsenthil2} demonstrated
that the dual theory Eq.~\ref{qed1} is the parent state of many
known strongly interacting boundary states of $3d$ TI, and these
boundary states can also be constructed using the original
physical Dirac fermion (electron). It is tempting to claim that
Eq.~\ref{qed1} is exactly dual to the free Dirac fermion (or
weakly interacting Dirac fermion), which is a very simple $(2+1)d$
conformal field theory (CFT). Recently a coupled wire construction
of the duality further supports this idea~\cite{mross}. In this
paper we will assume this duality is exact: namely Eq.~\ref{qed1}
is indeed a CFT in the infrared that is dual to the noninteracting
$(2+1)d$ Dirac fermion, and we will use this assumption to explore
other possible behaviors of the boundary.

The goal of this paper is to study the quantum phase transition
described by the following field theory: \beqn \mathcal{L} &=&
\bar{\psi}\gamma_\mu (\partial_\mu - i a_\mu) \psi +
\frac{1}{e^2}f_{\mu\nu}^2 \cr\cr &+& |(\partial_\mu - i k
a_\mu)\phi|^2 + r |\phi|^2 + g |\phi|^4, \label{L}\eeqn with
tuning parameter $r$, arbitrary integer $k$, Dirac fermion $\psi$
and complex scalar bosonic field $\phi$ which both couple to the
same $(2+1)d$ dynamical U(1) gauge field $a_\mu$. The boson $\phi$
can be viewed as the $4k$ fold vortex of the Fu-Kane
superconductor {\it bound} with another extra degree of freedom
(d.o.f). For even integer $k$, $\phi$ is the bound state of $4k$
vortex and an extra boson; while if $k$ is odd, $\phi$ must
contain an extra fermion that transforms in the same way as $\psi$
under $\mathcal{T}$, but neutral under the dynamical gauge field
$a_\mu$. The tuning parameter $r$ can be tuned by the mass gap of
this extra d.o.f.

Obviously this theory has two phases: when $r$ is sufficiently
large, $\phi$ is gapped, and based on our assumption the boundary
is described by Eq.~\ref{qed1}, and it is dual to a noninteracting
Dirac fermion; while when $r$ is negative and large, $\phi$
condenses, and it drives the boundary into a topological order
with {\it gapless} Dirac fermion $\psi$. We are interested in the
quantum phase transition between these two phases. Notice that
when $\phi$ condenses, $\psi$ is not automatically gapped, $i.e.$
there is no Yukawa type of coupling such as $\phi^\ast \psi^t
\gamma^0 \psi$ in the Lagrangian, which is forbidden by the gauge
symmetry for $k \neq 2$. It is easy to show that there is no other
obviously relevant couplings in Eq.~\ref{L} allowed by the gauge
symmetry.

Our goal is to calculate the scaling dimension of gauge invariant
order parameters and other universal quantities at the quantum
critical point $r = 0$. Let us take the limit $k \rightarrow +
\infty$ first. In this limit, the gauge field dynamics is
completely dominated by its coupling to the scalar field, and the
fermions will effectively decouple from the gauge field. More
precisely, the fermion decouples from the gauge field at the
energy scale below $k^2 e^2$. This effect becomes explicit after
we rescale $k a_\mu = \tilde{a}_\mu$. In this case the theory
becomes a standard bosonic QED with gauge field $\tilde{a}_\mu$,
and it is well-known that this theory is dual to a $3d$ XY
transition~\cite{halperindual,leedual}. We assume that we know
{\it everything} about the $3d$ XY transition, including all of
its critical exponents, the scaling dimension of all the composite
operators, the operator product expansion, and most importantly,
the universal boson conductivity
$\tilde{\sigma}$~\cite{fisher1,fisher2}, which we will take as a
dimensionless constant, assuming the boson carries charge$-1$. All
these information can be obtained by numerically studying the $3d$
XY transition only. For example, numerically the critical exponent
$\nu$ has been confirmed to be very close to (slightly larger
than) $2/3$~\cite{vicari}. The universal conductivity of the $3d$
XY transition has also been studied with various
methods~\cite{cond1,cond2,cond3,cond4}. Recent progresses based on
conformal bootstrap have determined the value of $\tilde{\sigma}$
very precisely~\cite{bootstrap}, which is highly consistent with
the numerical results~\cite{cond3,cond4}


{\it --- Scaling dimension of $\mathcal{T}$-breaking order
parameter}

Time-reversal symmetry $\mathcal{T}$ is the key symmetry that
protects the $3d$ TI. Let us compute the scaling dimension of the
time-reversal symmetry breaking order parameter $\bar{\psi}\psi =
\psi^\dagger \gamma^0 \psi$. In the large$-k$ limit, because
$\psi$ basically decouples from the gauge field (as we argued
above), the scaling dimension of $\bar{\psi}\psi$ is the same as
that of the free fermion $\Delta[\bar{\psi}\psi] = 2$. The
correction to this scaling dimension comes from the gauge
fluctuation $a_\mu$, thus we need to know the photon propagator
$G^a_{\mu\nu}$ in the large$-k$ limit.

In the large$-k$ limit, since this quantum phase transition
belongs to the $3d$ XY universality class, we assume that the
universal conductivity of the boson degrees of freedom which
carries the global U(1) symmetry of the $3d $ XY transition is a
known dimensionless constant $\tilde{\sigma}$. We know that in the
momentum-frequency space of the Euclidean space-time, the Kubo
formula gives us the following relation between the correlation
function of the boson current $J^\mu (p)$ and the universal
conductivity $\tilde{\sigma}$: \beqn \langle J_{\mu}(p) \
J_{\nu}(-p) \rangle = \tilde{\sigma} |p| \left( \delta_{\mu\nu} -
\frac{p_\mu p_\nu}{p^2} \right). \eeqn Then because the boson
current $J_\mu = \frac{k}{2\pi} \epsilon_{\mu\nu\rho}
\partial_\nu a_\rho$, the photon propagator at the quantum critical point
in the large$-k$ limit reads \beqn G^a_{\mu\nu} (p) =
\frac{\tilde{\sigma} (2\pi)^2}{k^2 |p|}\left( \delta_{\mu\nu} -
\frac{p_\mu p_\nu}{p^2} \right). \label{Ga} \eeqn Or in other
words Eq.~\ref{L} reduces to a bosonic QED in the large$-k$ limit,
and Eq.~\ref{Ga} describes the fully dressed gauge field
propagator. Throughout the paper we will choose the gauge
$\partial_\mu a_\mu = 0$.

The rest of the calculation is pretty standard: because the photon
propagator carries a factor $1/k^2$, a systematic expansion
controlled by small factor $1/k^2$ can be carried out. By
combining the vertex correction and the wave function
renormalization together, the scaling dimension of
$\bar{\psi}\psi$ at the $1/k^2$ order reads \beqn
\Delta[\bar{\psi}\psi] = 2 - \frac{16 \tilde{\sigma}}{3 k^2 }.
\eeqn A similar calculation of scaling dimension of fermion
bilinear operators of the standard QED$_3$ with large$-N$ flavors
of fermions can be found in
Ref.~\onlinecite{hermele2005,aliceafisher,xusachdev}. But let us
stress that in our case we only have one flavor of fermion and
boson field each.

{\it --- Scaling dimension of four-fermion interaction term}

\begin{figure}
\includegraphics[width=.35\textwidth]{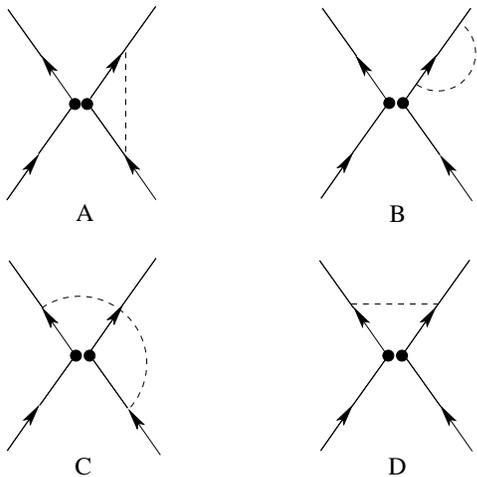}
\caption{The Feynman diagrams that will contribute to the the
scaling dimension of the four fermion interaction term at the
$1/k^2$ order.}\label{fd}
\end{figure}

A weak short range four-fermion interaction would be irrelevant
for a $(2+1)d$ gapless Dirac fermion. However, gauge fluctuation
potentially could change the scaling dimension of the four-fermion
interactions, and make them relevant. In our system Eq.~\ref{L},
because there is only one flavor of Dirac fermion $\psi$, there is
only one allowed four fermion interaction term without spatial
derivatives: \beqn g (\bar{\psi}\psi)^2 = - \frac{1}{3} g
(\bar{\psi} \gamma_\mu \psi)^2. \eeqn The scaling dimension of
this four fermion interaction term can again be calculated with a
$1/k^2$ expansion. All the Feynman diagrams that contribute at
this $1/k^2$ order are listed in Fig.~\ref{fd}. The final result
is \beqn \Delta[(\bar{\psi}\psi)^2] = 4 + \frac{16
\tilde{\sigma}}{3 k^2 }. \eeqn Thus the gauge fluctuation makes
the four-fermion interaction term even more irrelevant than it is
at the free Dirac fermion CFT. This calculation supports that
Eq.~\ref{L} describes a continuous quantum phase transition, since
the four fermion interaction is likely not rendered relevant by
gauge fluctuation for any $k$ at the quantum critical point $r=0$.

{\it --- Universal electrical conductivity}

As was pointed out in
Ref.~\onlinecite{maxashvin,wangsenthil1,wangsenthil2}, in
Eq.~\ref{L}, a $2\pi$ flux quantum of $a_\mu$ carries half
electric charge. Thus the physical electric current density at the
$2d$ surface reads $J^e_\mu = \frac{1}{4\pi}
\epsilon_{\mu\nu\rho}\partial_\nu a_\rho $. The electrical
conductivity $\sigma^e$ is encoded in the Euclidean space-time
correlation function of the current operator: \beqn \langle
J^e_\mu (p) \ J^e_\nu (-p) \rangle = \frac{\sigma^e}{e^2/\hbar}
|p| \left( \delta_{\mu\nu} - \frac{p_\mu p_\nu}{p^2} \right).
\label{cond} \eeqn When $\phi$ is gapped ($r
> 0$), the system is described by QED with $N=1$ flavor of Dirac
fermion $\psi$, which by our assumption is dual to a
noninteracting Dirac fermion which is not coupled to any dynamical
gauge field. Thus this phase with $r > 0$ is a semimetal with
universal electrical conductivity $\sigma^e =
\frac{1}{16}\frac{e^2}{\hbar}$. The quantum phase transition we
are studying is a transition from an electrical semimetal to an
electrical insulator, although the insulator phase is also
gapless.

Right at the quantum critical point, the electrical conductivity
must be a different universal value. Because we already know that
in the large$-k$ limit the photon $a_\mu$ propagator is given by
Eq.~\ref{Ga}, using the photon propagator, we can compute the
physical electric current-current correlation function: \beqn
\langle J^e_\mu (p) \ J^e_\nu (-p) \rangle  &=& \frac{1}{(4\pi)^2}
p^2 G^a_{\mu\nu} (p) \cr\cr &=& \frac{\tilde{\sigma}}{4k^2} |p|
\left( \delta_{\mu\nu} - \frac{p_\mu p_\nu}{p^2} \right). \eeqn
Comparing with Eq.~\ref{cond}, we conclude that in the large$-k$
limit the universal electrical conductivity at the quantum
critical point reads \beqn \sigma^e = \frac{\tilde{\sigma}}{4k^2}
\frac{e^2}{\hbar}. \label{condlargek} \eeqn The leading correction
to this value must be at the $1/k^4$ order, which comes from the
correction to the $a_\mu$ propagator from the Dirac fermion
$\psi$.

If the time-reversal symmetry $\mathcal{T}$ is broken at the $2d$
boundary, $i.e.$ the system develops a nonzero expectation value
of $\bar{\psi}\psi$, the Hall conductivity at the quantum critical
point $r = 0$ will also be at order $\sim
\frac{1}{k^4}\frac{e^2}{\hbar}$.

{\it --- Self-duality}

In this subsection we will see that Eq.~\ref{L} has a
(quasi-)self-dual structure. The duality transformation of the
second line of Eq.~\ref{L} is rather standard, it is simply the
particle-vortex duality: \beqn \mathcal{L}_b = |(\partial_\mu -
ib^{(1)}_\mu)\Phi|^2 + \tilde{r} |\Phi|^2 + \tilde{g} |\Phi|^4 +
\frac{i k}{2\pi} a \wedge d b^{(1)}, \label{d1}\eeqn where $\Phi$
can be viewed as the unit vortex field of $\phi$, and it is bound
with $2\pi k$ flux of $a_\mu$, because $\phi$ carries charge$-k$
under $a_\mu$. The duality of the first line of Eq.~\ref{L}
requires the newly developed (hypothesized) duality in
Ref.~\onlinecite{maxashvin,wangsenthil1,wangsenthil2}: \beqn
\mathcal{L}_f = \bar{\chi} \gamma_\mu (\partial_\mu - i
b^{(2)}_\mu) \chi + \frac{i}{4\pi} a \wedge db^{(2)}, \label{d2}
\eeqn where now $\chi$ transforms under time-reversal as
$\mathcal{T}: \chi \rightarrow i \sigma^y \chi$. If
$\mathcal{L}_b$ in Eq.~\ref{d1} is ignored, integrating out
$a_\mu$ in $\mathcal{L}_f$ will gap out $b^{(2)}_\mu$, thus
$\mathcal{L}_f$ only has a free Dirac fermion $\chi$ in the
infrared, which corresponds to the case studied in
Ref.~\onlinecite{maxashvin,wangsenthil1,wangsenthil2}

In our case, due to the existence of the bosonic matter field,
integrating out $a_\mu$ induces the following constraint: \beqn
b^{(2)}_\mu = - 2k b^{(1)}_\mu = - 2 k b_\mu. \eeqn Thus the final
dual theory reads \beqn \mathcal{L}_{dual} &=& \bar{\chi}
\gamma_\mu (\partial_\mu + i 2 k b_\mu ) \chi + \cdots \cr\cr &+&
|(\partial_\mu - ib_\mu)\Phi|^2 + \tilde{r} |\Phi|^2 + \tilde{g}
|\Phi|^4. \label{Ld} \eeqn Here $\tilde{r} \sim - r$: when
$\tilde{r} < 0$, $\Phi$ is condensed, which is dual to the
disordered phase of $\phi$, and low energy physics of this phase
is either described by a QED with $N = 1$ flavor of fermion
$\psi$, or a single gapless Dirac fermion $\chi$; When $\tilde{r}
> 0$, $\Phi$ is disordered, and the low energy physics of this
phase is described by either a QED with $N=1$ flavor of fermion
$\chi$, or a single gapless Dirac fermion $\psi$ (which is coupled
to a gapped discrete gauge field). The dual theory Eq.~\ref{Ld} is
very similar to the original theory Eq.~\ref{L}, the only
difference is that now it is the fermionic degree of freedom that
carries a large gauge charge.

Again, in the large$-k$ limit, Lagrangian Eq.~\ref{Ld} describes a
$3d$ XY transition, because after rescaling $k b_\mu =
\tilde{b}_\mu$, $\Phi$ is effectively neutral under
$\tilde{b}_\mu$ in the large$-k$ limit. Again, in the large$-k$
limit, the propagator of gauge field $b_\mu$ can be calculated
exactly, based on the observation that the fermion current
$J^\psi_{\mu} = \bar{\psi}\gamma_\mu \psi = \frac{1}{4\pi}
\epsilon_{\mu\nu\rho}\partial_\nu b^{(2)}_\mu = - \frac{k}{2\pi}
\epsilon_{\mu\nu\rho}\partial_\nu b_\mu$. In the large$-k$ limit
the correlation function of $J^\psi_\mu$ can be computed exactly
because in this limit $\psi$ decouples from $a_\mu$, and the
correlation function of $J^\psi$ in this limit is well-known:
\beqn \langle J^\psi_\mu(p) \ J^\psi_\nu(-p) \rangle =
\frac{1}{16} |p| \left( \delta_{\mu\nu} - \frac{p_\mu p_\nu}{p^2}
\right), \eeqn this implies that photon $b_\mu$ propagator in the
large$-k$ limit reads \beqn G^b_{\mu\nu} = \frac{\pi^2}{4 k^2 |p|}
\left( \delta_{\mu\nu} - \frac{p_\mu p_\nu}{p^2} \right). \eeqn

In this dual theory, operator $\bar{\chi}\chi$ breaks
time-reversal symmetry, and hence it can be identified as
$\bar{\psi}\psi$ in the original theory Eq.~\ref{L}~\cite{mross}.
Thus the scaling dimension of $\bar{\chi}\chi$ is also \beqn
\Delta[\bar{\chi}\chi] = 2 - \frac{16 \tilde{\sigma}}{3 k^2 }.
\eeqn

{\it --- Critical exponent}

We would also like to calculate the scaling dimension of the
tuning parameter $r$ in Eq.~\ref{L}, which is identified as
$\tilde{r}$ in the dual theory, thus the composite operator
$|\phi|^2$ is equivalent to $|\Phi|^2$.

\begin{figure}
\includegraphics[width=.39\textwidth]{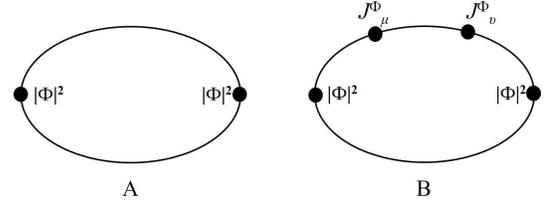}
\caption{The diagram on the right hand side is the correlation
function $\langle |\Phi^2|_{\vec{q}} J^{\Phi}_\mu(\vec{p})
J^{\Phi}_\nu ( - \vec{p}) |\Phi^2|_{-\vec{q}} \rangle$. When
$|\vec{p}| \gg |\vec{q}|$ it scales as the left hand side
correlation function times factor $\frac{1}{|p|^2}\left(
\delta_{\mu\nu} - \frac{p_\mu p_\nu}{p^2} \right)$.
}\label{bubble}
\end{figure}

To calculate the scaling dimension of $\tilde{r}$, one strategy is
to expand Eq.~\ref{Ld} at the Gaussian fixed point of $\Phi$ and
perform a combined $\epsilon = 4 - D$ and $1/k^2$ expansion.
Although this calculation is straightforward, we hope to expand
everything at the $3d$ XY fixed point (which we assume to know
everything about) in the large$-k$ limit. In order to carry out
the renormalization group (RG) calculation, we make use of the
operator product expansion (OPE) in the momentum space: \beqn &&
\left( \frac{1}{2} \tilde{r}|\Phi|^2 J^{\Phi}_\mu(\vec{p})
J^{\Phi}_\nu ( - \vec{p}) \right) G^b_{\mu\nu}(\vec{p}) \cr\cr
\sim && \left( \frac{1}{2} \tilde{r}|\Phi|^2 \frac{C}{p^2} \left(
\delta_{\mu\nu} - \frac{p_\mu p_\nu}{p^2} \right) \right)
G^b_{\mu\nu}(\vec{p}) \cr\cr = && \left( \tilde{r}|\Phi|^2
\frac{C}{|p|^3} \right) \frac{\pi^2}{4k^2}. \label{ope} \eeqn
$J_\mu^\Phi(\vec{p})$ is the current operator of field $\Phi$ in
Eq.~\ref{Ld}. The meaning of this OPE is that, when the momentum
$\vec{p}$ of $J_\mu^{\Phi}$ and the photon propagator is much
larger than the momentum of $|\Phi|^2$, the correlation function
between the composite operator $|\Phi|^2 J^{\Phi}_\mu(\vec{p})
J^{\Phi}_\nu ( - \vec{p})$ and another operator can be
approximated by the correlation between $|\Phi|^2
\frac{C}{|p|^2}\left( \delta_{\mu\nu} - \frac{p_\mu p_\nu}{p^2}
\right)$ and that operator. We have checked this OPE by comparing
the two Feynman diagrams in Fig.~\ref{bubble}, and the correlation
function $\langle |\Phi^2|_{\vec{q}} J^{\Phi}_\mu(\vec{p})
J^{\Phi}_\nu ( - \vec{p}) |\Phi^2|_{-\vec{q}} \rangle$ indeed
scales as $\sim \langle |\Phi^2|_{\vec{q}} \ |\Phi^2|_{-\vec{q}}
\rangle \frac{1}{|p|^2}\left( \delta_{\mu\nu} - \frac{p_\mu
p_\nu}{p^2} \right) $ when $|\vec{p}| \gg |\vec{q}|$.

In this OPE, the dimensionless number $C$ only depends on the $3d$
XY universality class, and as we stated we assume that it can be
determined by studying the OPE of the $3d$ XY transition only,
through for instance the $1/N$ expansion as in
Ref.~\onlinecite{william}. Although the dimensionless number $C$
is yet to determine, the $1/|\vec{p}|^2$ scaling of this OPE is
known, because the scaling dimension of the boson current
$J^{\Phi}_\mu$ is $\Delta[J^{\Phi}_\mu] = 2$, which at the $3d$ XY
fixed point is unrenormalized compared with the free boson theory
because it is a conserved current.

After the standard momentum shell RG calculation, $i.e.$
integrating out the degrees of freedom with momentum $\vec{p}$
between $ b \Lambda < |p| < \Lambda $, the OPE above will
contribute a correction to $\tilde{r} |\Phi|^2$ that is
proportional to $\ln (1/b)$. Now we can conclude that the RG
equation for $\tilde{r}$ to the $1/k^2$ order reads \beqn \frac{d
\tilde{r}}{ d\ln (1/b)} = \left( \Delta_{xy} + \frac{C}{8k^2}
\right) \tilde{r}, \eeqn which determines the scaling dimension of
$\tilde{r}$. Here $\Delta_{xy}$ is the scaling dimension of
$\tilde{r}$ at the $3d$ XY universality class, which is very close
to $3/2$~\cite{vicari}.

{\it --- Summary}

In this work we did our best to study the the quantum phase
transition described in Eq.~\ref{L}, with its dual Lagrangian
described by Eq.~\ref{Ld}. The self-dual nature of this transition
allows us to calculate many quantities in a controlled expansion
with $1/k^2$. But it is possible that, with small enough $k$, the
transition becomes first order.

The same techniques used in this work can be applied to other
field theories as well. For instance QED$_3$ with two flavors of
Dirac fermions, and one flavor of fermion carries gauge
charge$-1$, while the other flavor of fermion carries a much
larger gauge charge$-k$. A similar $1/k^2$ expansion can also be
applied to this theory as well.

Ref.~\onlinecite{mirror} has applied the mirror
symmetry~\cite{mirror1,mirror2,mirror3} (duality between
supersymmetric field theories) to the half-filled Landau
level~\cite{son2015}, which is a system closely related to the
boundary of $3d$ TI~\cite{maxashvin,wangsenthil1,wangsenthil2}.
Previous study~\cite{mirrorsubir} also indicates that the mirror
symmetry is related to the ``deconfined
QCP"~\cite{deconfine1,deconfine2}. We suspect the QCP discussed in
this paper may also have an interesting supersymmetric version. We
will leave this to future study.

The authors are supported by the David and Lucile Packard
Foundation and NSF Grant No. DMR-1151208. The authors are grateful
to Leon Balents, Matthew Fisher, William Witczak-Krempa, Ashvin
Vishwanath for very helpful discussions.

\bibliography{QCP}

\end{document}